\documentclass[12pt]{article}

\input psfig

\title{
       An Exact Solution to the Time-dependent Schr\"{o}dinger Equation
       for a Model One-dimensional Potential
      }

\author{
         { Athanasios N. Petridis,          } 
         { Lawrence P. Staunton,            }
         { Jon Vermedahl\footnote{Now at the University of Minnesota, 
                                  Minneapolis, MN 55455, USA.} } \\
         {\em Department of Physics and Astronomy,}
         {\em Drake University                    } \\
         {\em Des Moines, IA 50311, USA           } \\
         { and                                    } \\
         { Marshall Luban                   } \\
         {\em Ames Laboratory and Department of Physics and Astronomy, } \\
         {\em Iowa State University, Ames, IA 50011, USA} 
       }

\begin{document}

\maketitle

\begin{abstract}
Analytical solutions to the time-dependent Shr\"{o}dinger
equation in one dimension are developed for time-independent 
potentials, one consisting of an infinite wall and a repulsive 
delta function. An exact solution is
obtained by means of a convolution of time-independent 
solutions spanning the given Hilbert space with appropriately 
chosen spectral functions. Square-integrability and 
the boundary conditions are satisfied. The probability for 
the particle to be found inside the potential well is 
calculated and shown to exhibit non-exponential decay decreasing
at large times as $t^{-3}$. The result is generalized for
all square-integrable solutions to this problem.

\end{abstract}

\vspace{0.2in}


\section{Introduction}

Time-dependent quantum mechanics problems are usually addressed
using time-dependent perturbation theory, adiabatic or sudden 
approximations as well as several numerical techniques. It is
highly desirable, however, to obtain exact analytical solutions
to given problems, especially in cases when the approximate
methods may be inadequate to describe detailed aspects of the
solutions or when numerical treatment does not explicitly 
reveal their mathematical properties at very large times.
Recently, there has been increasing interest in obtaining exact 
analytical solutions to the time-dependent Schr\"{o}dinger
equation since they can be used to study certain physical 
systems such as quantum dots, Bose-Einstein condensates, unstable
composite particles and many others. 
Burrows and Cohen \cite{Burrows} have developed exact solutions
for a double-well quasi-harmonic potential model with a time-dependent
dipole field. Cavalcanti, Giacconi and Soldati \cite{Cavalcanti}
have solved the problem of decay from a point-like potential well
in the presence of a uniform field and have indicated that, due to
an infinitely large number of resonances, there may be deviations from
the naively expected exponential time-dependence of the survival 
probability. 

The equivalence of exponential decay of a given 
energy eigenstate with Fermi's golden rule when the final density 
of states is energy-independent and with the Breit-Wigner resonance 
curve has been long known in quantum mechanics \cite{Wigner,Sakurai}. 
Dullemond \cite{Dullemond} has verified this
fact in a simple but exactly solvable model and found
that if final-state energy-dependence is introduced into this model
a non-exponential decay pattern will dominate at large times. 
Petridis {\em et al.,} \cite{Petridis} have studied a variety of 
systems in which the initial wave function is mostly or entirely set 
in a finite potential well and have observed rich behaviour, including 
non-exponential decay into the continuum. Non-exponentiality
for monotonically decreasing survival probabilities at large
times, though, can be strictly proven only if exact analytical 
solutions are obtained. Specific systems that may exhibit 
non-exponential decay include systems with non-local 
interactions \cite{Shirokov}, certain closed many-body systems \cite{Flambaum}, 
quasi-particles in quantum dots \cite{Silvestrov}, polarons \cite{Acardi}, 
and non-extensive systems \cite{Wilk}. 

In this article a method for developing analytical solutions
to the time-dependent Schr\"{o}dinger equation is presented.
The method is applied to a time-independent potential, 
consisting of an infinite wall and a repulsive delta function. 
In addition to the mathematical interest
that this potential exhibits it is also applicable to a
variety of quantum systems undergoing decay. A wavefunction
that is an exact solution subject to the boundary conditions
is obtained and used to analytically calculate the probability 
for finding the particle inside the potential well at any time.   
A generalization of the asymptotic time behaviour to all 
square-integrable wavefunctions is obtained. 

\section{The proposed method}

The method consists of the following steps: (a) The time-independent
solutions to Schr\"{o}dinger equation are found subject to
the boundary conditions of the problem. These are stationary solutions,
that is energy eigenfunctions, that span the Hilbert space of the
given Hamiltonian. (b) Since any finite or infinite, discrete
or continuous linear combination of the stationary solutions (base 
functions) is also a solution belonging to the given Hilbert space, 
as long as it is square-integrable, exact analytical solutions can be 
developed by a convolution of the eigenfunctions with energy-dependent 
spectral functions multiplied by the standard oscillatory time-dependence
of the stationary states. It is, obviously, necessary that the convolution 
integral over the energy converge. This convergence as well as the 
square-integrability (normalizability) of the resulting wave function are 
verified. (c) The survival probability, i.e., the probability for finding the
particle inside the potential well is calculated and its properties
are studied analytically.

\section{Infinite wall and delta-function potential}

The problem to be considered is defined by the
one-dimensional repulsive potential,
\begin{equation}
V(x) = \infty,  \:\:\: -\infty \le x \le 0  \: \: {\rm and} \: \: 
V(x) = V_{0} \: \delta(x-L),  \:\:\: 0 < x \le \infty, 
\label{wallanddeltapot}
\end{equation}
with $L >0$ and $V_0 > 0$. The steps outlined in the previous section 
are followed. 

(a) The solutions to the time-independent Schr\"{o}dinger equation,
\begin{equation}
- \frac{1}{2} \frac{d^{2}\Psi_{E}(x)}{dx^{2}} + V(x) \Psi_{E}(x) = E \: \Psi_{E}(x), 
\label{timeindependentSeq}
\end{equation}
(with particle mass $m = 1$, $\hbar = 1$ and $E \ge 0$) are,
\begin{eqnarray}
\Psi_{E}^{(0)}(x) & = & 0, \:\: -\infty \le x \le 0 \:\: {\rm (region \: "0")}, \\
\Psi_{E}^{(I)}(x) & = & C_{1} \sin{(px)}, \:\: 0 \le x \le L \:\: {\rm (region \: "I")}, \\
\Psi_{E}^{(II)}(x) & = & C_{2} \sin{(px)} + C_{3} \cos{(px)},  \:\: L \le x \le \infty
                \:\: {\rm (region \: "II")},  
\label{wallanddeltasol}
\end{eqnarray}
where $p = \sqrt{2E}$ and $C_{1,2,3}$ are constants in $x$. These functions obey the boundary
conditions
\begin{eqnarray}
\Psi_{E}^{(I)}(L) & = & \Psi_{E}^{(II)}(L), \\
\frac{d \Psi_{E}^{(I)}}{dx}(L) & - & \frac{d \Psi_{E}^{(II)}}{dx}(L)  =  
                                    2 V_{0} \: \Psi_{E}^{(I)}(L),
\end{eqnarray}
while the boundary conditions at $x = 0$ are automatically satisfied. The solution is not
required to vanish at infinity since functions that do not vanish at large $x$ can still 
be solutions to the time-dependent problem. Selecting $C_1$ as the overall normalization
constant the boundary conditions at $x=L$ yield
\begin{eqnarray}
C_2 & = & C_1 \left [1 + \left (\frac{2V_0}{p} \right ) \: \sin{(pL)} \: \cos{(pL)} \right ], \\
C_3 & = & - C_1 \left ( \frac{2V_0}{p} \right ) \: \sin^2{(pL)},
\end{eqnarray}
rendering $C_2$ and $C_3$ functions of the energy. The choice of $C_2$ or $C_3$ as
the normalization constant would introduce an energy-dependence in $C_1$ and would
effectively amount to different spectral functions. 

The obtained linearly independent energy eigenfunctions are orthogonal under the
inner product
\begin{equation}
(\psi_1, \psi_2) = \int_{0}^{L} \psi^{*}_1(x) \: \psi_2(x) \: dx + 
              \lim_{\epsilon \rightarrow 0} 
              \int_{L}^{\infty} e^{-\epsilon (x - L)} \: \psi^{*}_1(x) \: \psi_2(x) \: dx,  
\label{innerprod}
\end{equation}
with all wavefunctions in the defined Hilbert space identically vanishing for $x \leq 0$.
The orthogonality relation is
\begin{equation}
(\Psi_{E}, \Psi_{E^{'}}) = w(E) \: \delta (p - p^{'}),
\label{orthogonal}
\end{equation}
where $p = \sqrt{2E}$, $p^{'} = \sqrt{2E^{'}}$ and
\begin{eqnarray}
w(E) & = & \frac{\pi}{2} \left [ |C_2(E)|^2 + |C_3(E)|^2 \right ] \\ \nonumber
     & = & |C_1|^2 \frac{\pi}{2p^2} 
       \left [ p^2 + 2 V_0^2 - 2 V_0^2 \cos{(2pL)} + 2pV_0 \sin{(2pL)} \right ]. 
\label{wfunction}
\end{eqnarray} 
The Dirac $\delta$-function representation used is
\begin{equation}
\delta(x) = \lim_{\epsilon \rightarrow 0} \frac{1}{\pi} \: \frac{\epsilon}{x^2 + \epsilon^2}.
\end{equation}

(b) The solution to the time-dependent Schr\"{o}dinger equation,
\begin{equation}
-i \frac{\partial \psi (x,t)}{\partial t} = 
- \frac{1}{2} \frac{\partial^{2} \psi(x,t)}{\partial x^{2}} + V(x) \: \psi(x,t),
\end{equation}
can be written as the energy-convolution integral,
\begin{equation}
\psi(x,t) = \int_{0}^{\infty} \phi(E) \Psi_{E}(x) e^{-iEt} \: dE,
\label{timedepsol}
\end{equation}
with $\phi(E)$ a spectral function such that
this integral is convergent for all $x$ and all $t$ and the resulting wavefunction
is square-integrable. Square-integrability of $\psi(x,t)$ also requires $E$ to be real.
The overall normalization constant is, then, calculated from
\begin{equation}
\int_{0}^{\infty} \psi^{*}(x,t) \psi(x,t) \: dx = 1.
\label{normalization}
\end{equation}

The choice of spectral function
\begin{equation}
\phi(E) = e^{-K^2 E}, 
\label{spectralfunction}
\end{equation}
with $K$ a positive constant, produces a convergent energy-convolution integral
and a wavefunction that is square-integrable even without the presence of
the convergence factor that appears in eq.~(\ref{innerprod}). These integrals
can be evaluated analytically and in closed form.
The time-dependent solution is, then,
\begin{eqnarray}
\psi^{(0)}(x,t) & = & 0, \\
\psi^{(I)}(x,t) & = & C_1 \: \sqrt{\frac{\pi}{2}}\: x \: e^{-\frac{x^2}{2(K^2 + it)}} 
                        \: (K^2 + it)^{-3/2} ,
\label{deltasolI} 
\end{eqnarray}
\begin{eqnarray}
\psi^{(II)}(x,t) & = & C_1 \: \sqrt{\frac{\pi}{2}} \: e^{-\frac{2L^2 - 2Lx + x^2}{K^2 + it}} 
                     \: (K^2 + it)^{-3/2} \nonumber \\
                 &   & \left [ e^{\frac{x^2}{2(K^2 + it)}} (K^2 + it) V_0
                           +  e^{\frac{(-2L+x)^2}{2(K^2+ it)}} (-K^2 V_0 - it V_0 + x)
                       \right ],
\label{deltasolII}
\end{eqnarray}
where
\begin{equation}
C_1 =  \left [ \frac{\pi^{3/2}}{8K^3} +
             \frac{e^{-\frac{L^2}{K^2}} L \pi^{3/2} V_0}{2K^3}
           + \frac{\pi^{3/2} V_0^2}{2K} 
           - \frac{e^{-\frac{L^2}{K^2}} \pi^{3/2} V_0^2}{2K}
\:\: \right ]^{-1/2}
\label{deltasolnorm}
\end{equation}
is the overall normalization factor obtained by means of eq.~(\ref{normalization}).

(c) The probability density $\rho = \psi^{*}(x,t) \psi(x,t)$ can be calculated for the interior
(region I) and the exterior (region II) of the potential well. It is presented in Fig. 1 at
six times starting from $t = 0$, in increasing order. The initial wave function is not entirely
localized inside the well. As time progresses the wavefunction spreads and tunnels through the
potential barrier in both directions. The interference of the wave that propagates outwards 
through the barrier and the wave that is outside creates the observed ripples. Inside the well
there are no ripples because the wavefunction is forced to be odd in $x$, having a node at $x=0$.
The centroid of the probability density in region II at $t=0$ is always located at $2L$, 
regardless the value of $K$. 
 
\begin{figure}
\centerline{\psfig{figure=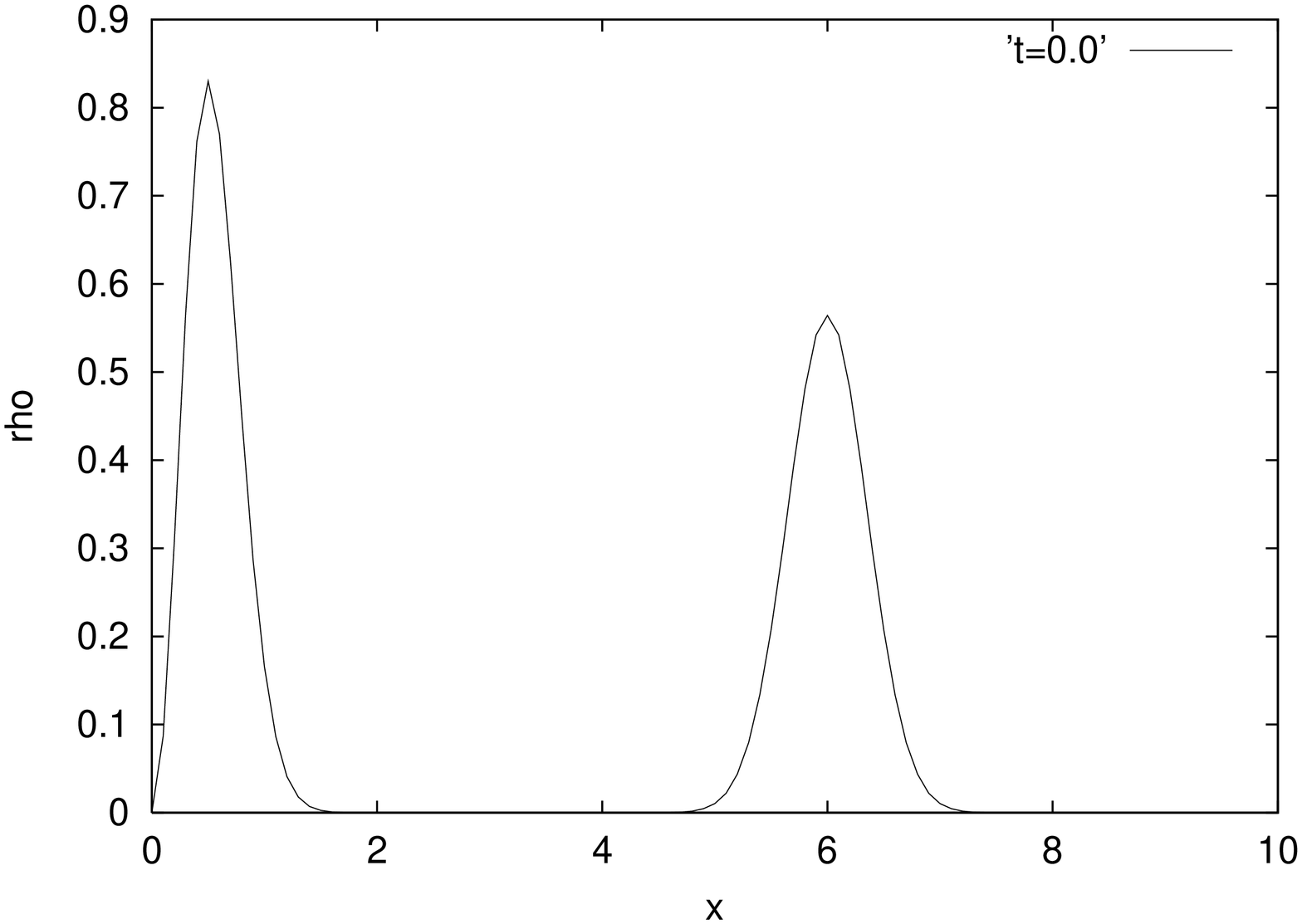,height=2.2in,angle=270}
            \psfig{figure=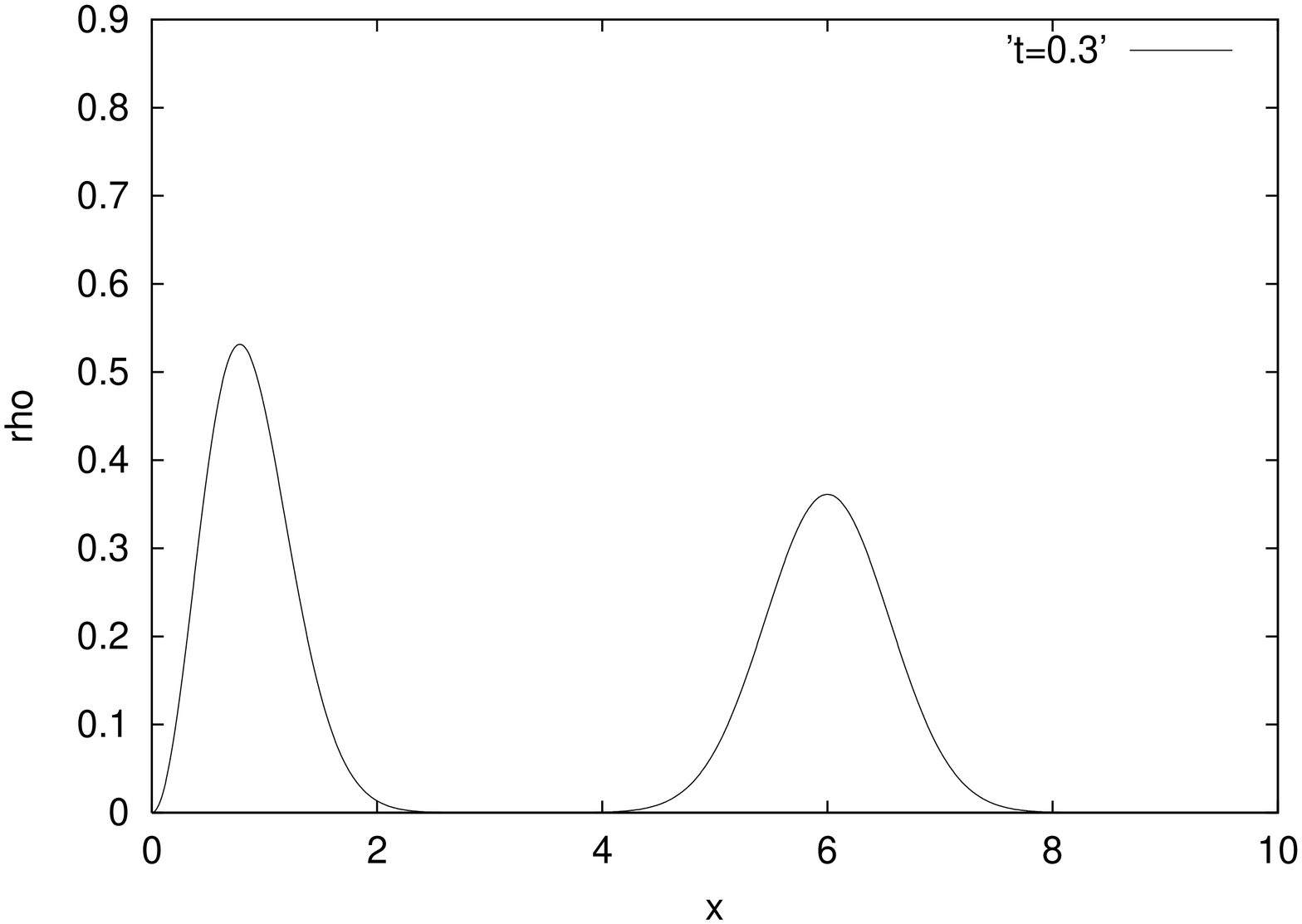,height=2.2in,angle=270}}
\centerline{\psfig{figure=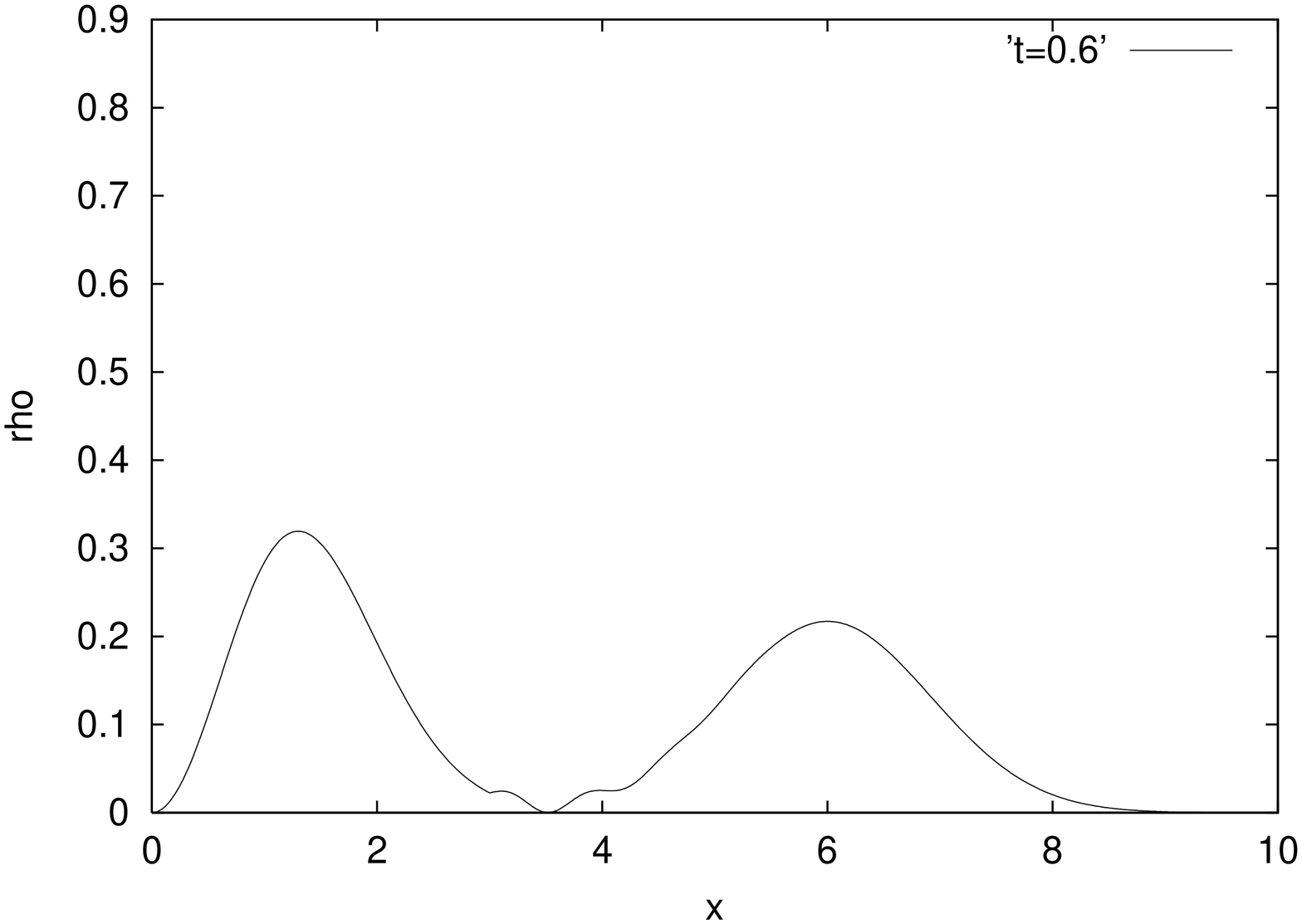,height=2.2in,angle=270}
            \psfig{figure=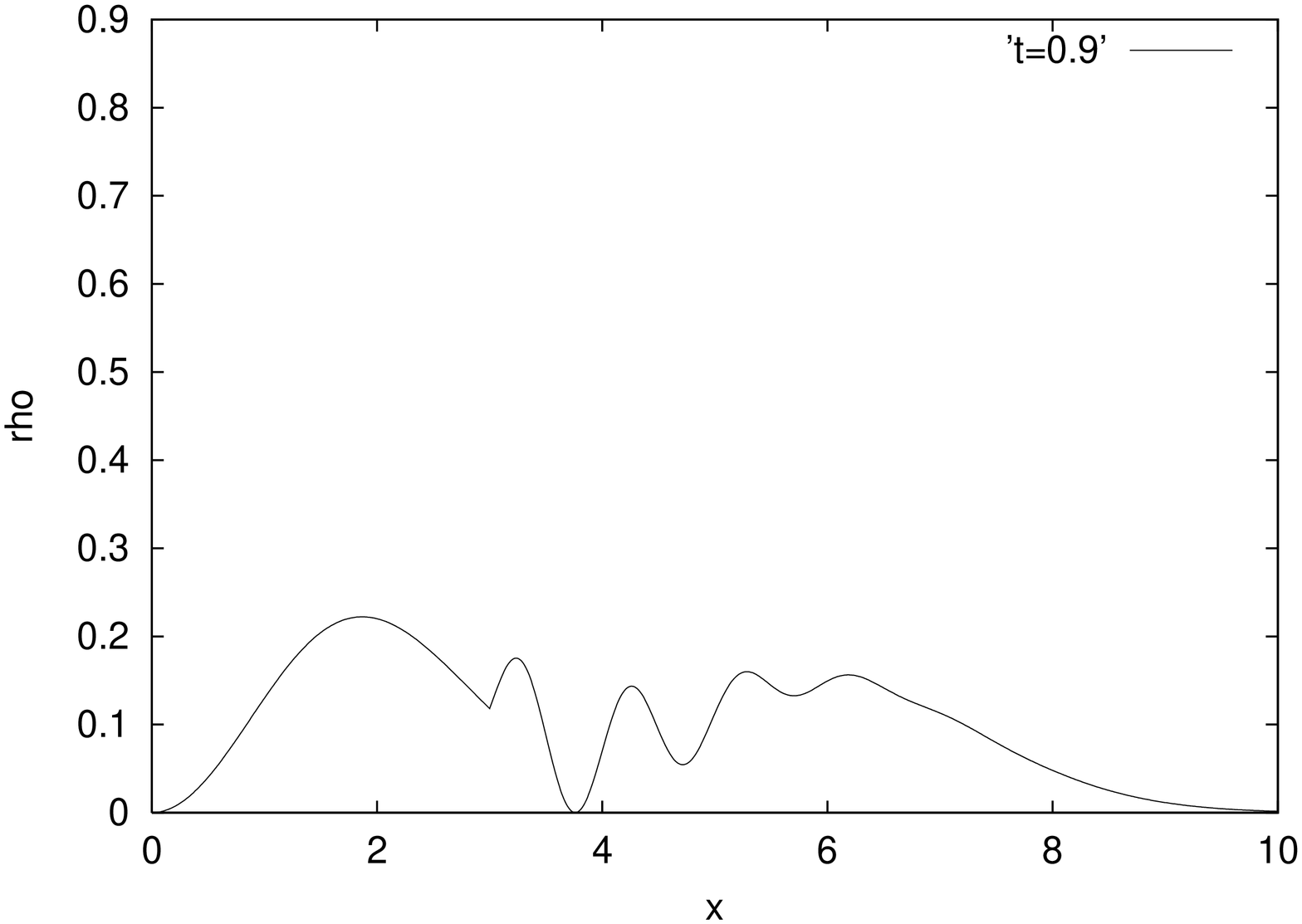,height=2.2in,angle=270}}
\centerline{\psfig{figure=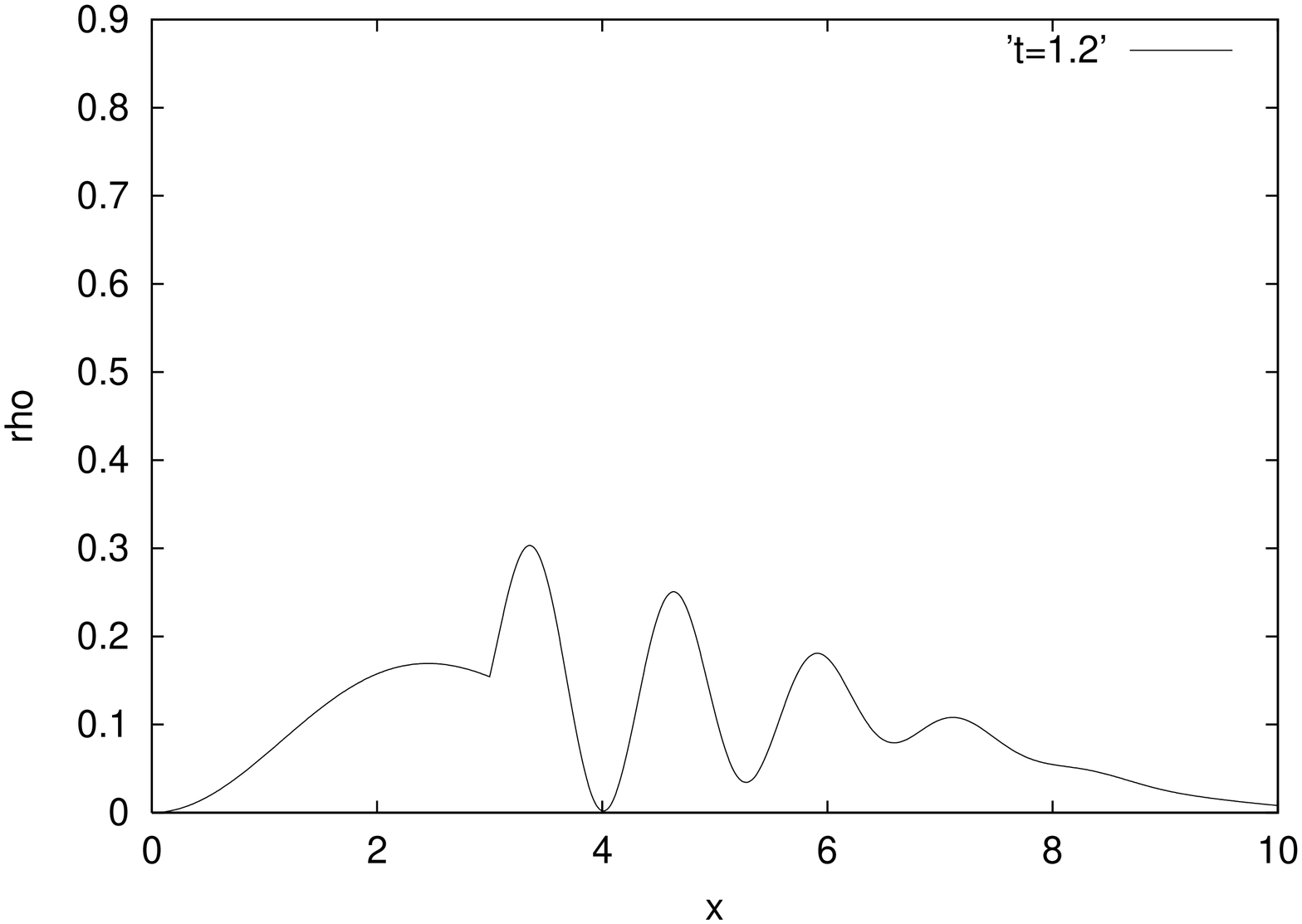,height=2.2in,angle=270}
            \psfig{figure=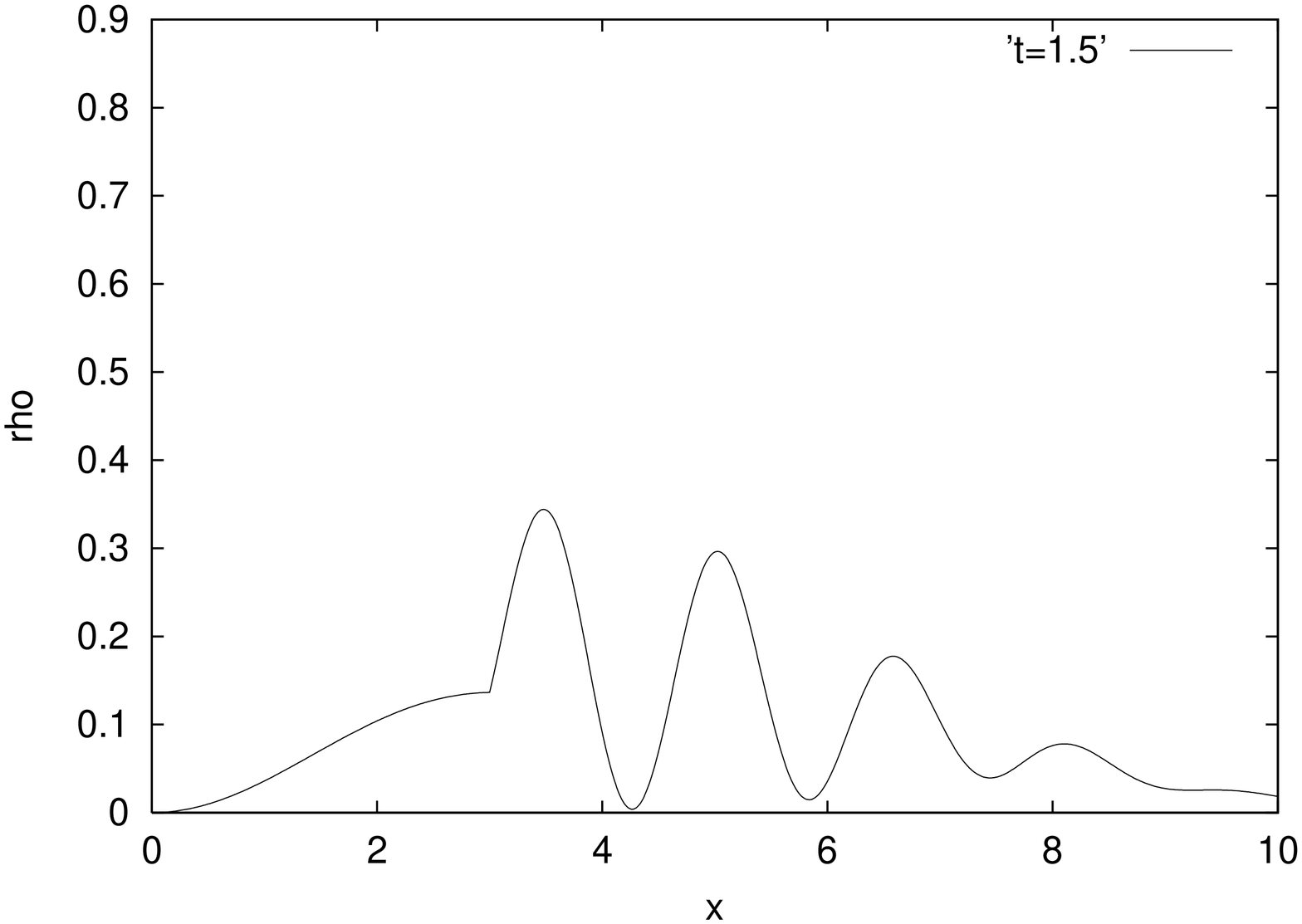,height=2.2in,angle=270}}
\caption{The probability density for a potential consisting of an infinite 
wall and a repulsive delta function and using a spectral function that is exponential in the 
energy at six increasing times (from the upper left to the lower right panel, 
$t=0.0,\:0.3,\:0.6,\:0.9,\:1.2,\:1.5$). In this plot $L = 3$, $V_0 = 1$ and $K = 1/2$. }
\label{deltaandwallrhofig}
\end{figure}

The survival probability is, then, calculated as
\begin{equation}
P_{in}(t) = \int_{0}^{L} \psi^{*}(x,t) \psi(x,t) \: dx.
\label{survivalprob}
\end{equation}
This yields the closed-form result
\begin{equation}
P_{in}(t) = C_1^2 \: \left [
 \frac{\pi^{3/2}}{8 K^3} \: \rm{Erf} \left ( \frac{KL}{\sqrt{K^4 + t^2}} \right )
       \: - \: \frac{2 \pi KL}{8K^3 \sqrt{K^4 + t^2}} e^{-\frac{K^2 L^2}{K^4 + t^2}} \right ].
\label{deltaandwallpin}
\end{equation}
A plot of the survival probability versus time is given in Fig. 2. $P_{in}(0)$ is controlled
by $K$ and approaches an upper limit as $L$ increases at fixed $K$. 
For example, this limit is equal to 0.9615 for $K=0.1$, 0.5 at $K=0.5$ and
0.1468 at $K=1.2$. It decreases as $K$ increases, i.e., as the momentum spectrum becomes sharper. 
On the other hand the decay becomes slower as $K$ increases. 
The expansion of $P_{in}$ in inverse powers of time includes only odd terms with alternating 
sings. At large times the leading term, that has a positive sign, is proportional to $t^{-3}$, 
a clearly non-exponential behaviour.

\begin{figure}
\centerline{\psfig{figure=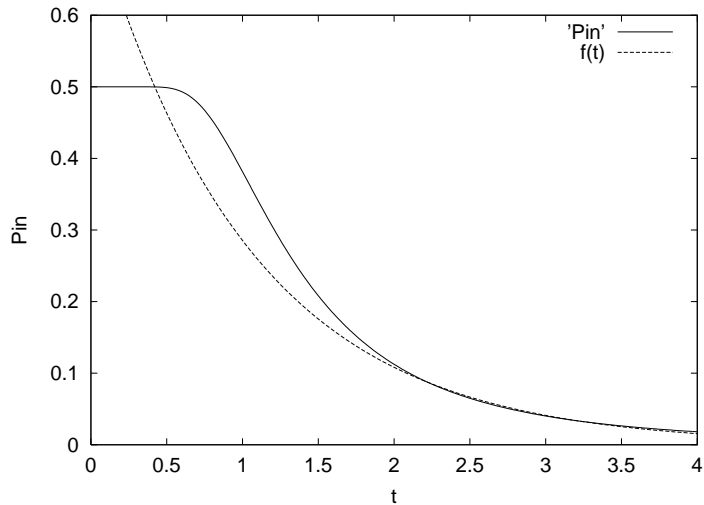,height=2.7in,angle=270}}
\caption{The survival probability for a potential consisting of an infinite wall and
a repulsive delta function and using a spectral function that is exponential in the energy
versus time (solid line). In this plot $L = 3$, $V_0 = 1$ and $K = 1/2$. The dashed line
represents the exponentially decaying function, $f(t) = a\: {\rm exp}(-b \: t)$,
fitted to data points, calculated from the actual solution, in the range $t=2$ to 4. 
The $\chi^2$ per degree of freedom is of order $10^{-6}$.}
\label{deltaandwallpinfig}
\end{figure}

\section{Corrections to the exponential decay law}

The law governing the decay of physical systems is typically assumed to be a simple
exponential time-dependence of the number $N(t)$ of the systems that have not decayed
until time $t$, i.e, $N(t) = N(0) \: {\rm exp}(-\lambda t)$, where $\lambda$ is the decay 
constant. As mentioned earlier this simple law is consistent with the Breit-Wigner curve and 
Fermi's golden rule if the final density of states is energy independent. It refers to the 
survival probability of a given initial energy eigenstate. In the system studied here the 
initial state is not an eigenstate of the energy. If a very large number of systems is assumed 
to be initially described by $\psi(x,0)$ and a system is said to have decayed if the particle 
has exited the potential well, then the number of surviving systems is proportional to the 
probability $P_{in}$, i.e.,
\begin{equation}
\frac{N(t)}{N(0)} = \frac{P_{in}(t)}{P_{in}(0)}.
\label{nequation}
\end{equation}  
The differential decay law is
\begin{equation}
dN = -\lambda(t) N(t) \: dt,
\end{equation}
where, $\lambda$ is, in general, dependent on time. Substitution from
eq.~(\ref{nequation}) gives
\begin{equation}
\lambda(t) = -\frac{1}{P_{in}}\frac{dP_{in}}{dt} = -\frac{d}{dt}[{\rm ln}(P_{in}(t))].
\end{equation}
In the case studied, eq.~(\ref{deltaandwallpin}) yields
\begin{equation}
\lambda(t) = \frac{4 e^{-z^2} z^3 t}{(K^4 + t^2) [-2 z e^{-z^2} + \sqrt{\pi} 
          \: {\rm Erf}(z) ] },
\end{equation}
where $z = KL/\sqrt{K^4 + t^2}$. This function is plotted versus time in Fig. 3.

\begin{figure}
\centerline{\psfig{figure=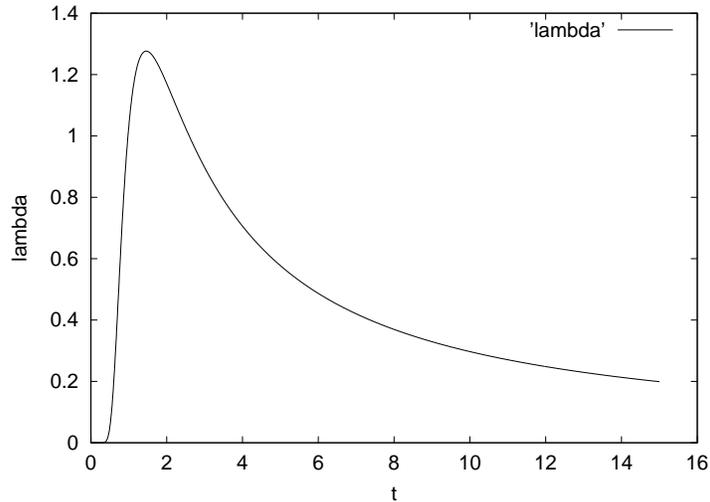,height=2.7in,angle=270}}
\caption{The decay parametre $\lambda$ for a potential consisting of an infinite wall and
a repulsive delta function and using a spectral function that is exponential in the energy
versus time. In this plot $L = 3$ and $K = 1/2$. There is no dependence on $V_0$.}
\label{lambdafig}
\end{figure}

The decay parameter $\lambda$ peaks later in time and has a smaller maximum value as $K$
or $L$ increase but does not depend on $V_0$.
The expansion of $\lambda$ in inverse powers of time includes only odd terms with alternating
sings. At large times the leading term, that has a positive sign, is proportional to $t^{-1}$, 
affirming the non-exponential behaviour. At very large times the change of $\lambda$ with time
is rather slow. A fit to $P_{in}$ at large times with an exponential curve in a finite time 
interval (as it is done in experiments) gives a very small value of
$\chi^2$ per degree of freedom (of order $10^{-6}$) so that the distinction between $P_{in}$ 
at large times and a simple exponential decay function is numerically minute (Fig. 2). 

\section{Generalization}

Exact, closed-form, analytical solutions to the time-dependent Schr\"{o}dinger equation for
the potential consisting of an infinite wall and a repulsive delta function can be and have been
obtained for other spectral function choices. For example, the choice 
\begin{equation}
\phi(E) = - \frac{i \left [1 - \cos{\left (\frac{L}{2} \: 
                    \sqrt{2E} \right )} \right ]}{2E \sqrt{\pi L} }
\end{equation} 
yields a square-integrable wavefunction.
In the absence of the delta function at $x = L$ this would produce an effectively square density 
pulse at $t = 0$ located between $x = 0$ and $x = L/2$. Due to the actual boundary conditions 
at $x = L$ this spectral function also produces a cusp centered at $x = 2L$. 
The survival probability is readily expressible in terms of Fresnel sine and
cosine integrals. Its asymptotic large time behaviour is still $t^{-3}$. 

A question that naturally arises at this point is whether the asymptotic time behaviour
can be generalized to other possible solutions to this problem. There is a one-to-one 
correspondence between spectral functions and square-integrable wavefunctions. This can be 
seen upon projecting the wavefunction at $t = 0$ on an energy eigenfunction and employing 
the orthogonality condition of eq.~(\ref{orthogonal}):
\begin{equation}
\phi(E) = \frac{1}{w(E)} \int_{0}^{\infty} \Psi_{E}^{*}(x) \Psi(x,0) \: dx. 
\end{equation}
Given an initial wavefunction the corresponding spectral function can, in principle,
be constructed. Schr\"{o}dinger's time-dependent equation then produces the wavefunction
at any later (or earlier) time.  

Convergence of the energy convolution integral in region (II) requires that the spectral 
function be finite at $E \rightarrow 0$. In addition, in order for $\Psi(x,t)$ to be 
square-integrable, $\phi(E)$ must vanish at large energies. This requirement can be made 
precise by inserting eq.~(\ref{timedepsol}) into eq.~(\ref{normalization}) and applying
eq.~(\ref{orthogonal}) to obtain
\begin{equation}
\int_{0}^{\infty} w(E) |\phi(E)|^2 \:dE = 1. 
\end{equation}
Inspection of the function $w(E)$, given in eq.~(12), leads to the conclusion
that $|\phi(E)|$ must vanish at infinity faster than $1/\sqrt{E}$ due to 
a constant term in $w(E)$.

Assuming that $\phi(E)$ satisfies the convergence conditions, its contribution to the
energy convolution integral giving $\Psi^{(I)}(x,t)$ comes mostly from low
energies. Then at any $x$ in region (I) the wavefunction can be approximated as
\begin{equation}
\Psi^{(I)}(x,t) \approx C_1 \phi(0) \int_{0}^{E_{max}(t)} \sqrt{2E} \: x \: e^{-iEt} \: dE.
\end{equation}
The upper limit of the integration is chosen as follows:
the factor $\exp{(-iEt)}$ oscillates more rapidly with the energy as $t$ increases.
At very large times these oscillations eventually lead to a vanishing contribution to
the integral. Therefore, the integral can be cut off at a point $E_{max}(t)$
whose first order term in the expansion in powers of $1/t$ is $y_{max}/t$, where
$y_{max}$ is constant in $t$.  
At low energies $\phi(E)$ is replaced by its (finite) value at $E = 0$ and
the sine function is substituted by its argument at a given $x$. 
Then, the variable change $y=Et$ yields
\begin{equation}
\psi^{(I)} \approx C_1 \phi(0) \: x \: 
           t^{-3/2} \int_{0}^{y_{max}} \sqrt{2y} \: e^{-iy} \: dy. 
\end{equation} 
For small $y_{max}$ the integral is  
approximately $\sqrt{2} \: [(2/3) \: y_{max}^{3/2} - i \: (2/5) \: y_{max}^{5/2}]$. 
The wavefunction in region (I) is to the first non-vanishing order
\begin{equation}
\Psi^{(I)}(x,t) \approx C_1 \phi(0) \: x \: M \: t^{-3/2},
\label{approxpsi}
\end{equation}    
where $M$ is a constant and the survival probability decreases with time as $t^{-3}$. Therefore, 
in order for the wavefunction to be square-integrable, the spectral function must be finite at 
$E \rightarrow 0$ and decrease at large $E$ faster than $1/\sqrt{E}$ and then necessarily 
the survival probability asymptotically decreases as the inverse cube of time. 

This argument can be extended to any finite value of $x$ including region (II)
since the coefficients $C_2$ and $C_3$ are at most of $O(1)$ for small $E$. Therefore, the 
integral of the probability density over any finite range of $x$ is finite (even without
the convergence factor present in eq.~(\ref{innerprod})) and it decreases asymptotically 
as $t^{-3}$. 

The constant $M$ in eq.~(\ref{approxpsi}) can be exactly evaluated if $\phi(E)$ decreases
at large $E$ faster than $1/E$. Then if $\phi(E)$ is analytic in the fourth quadrant of the
complex $E$-plane the contour integral of $\phi(E) \: \sin{(x\sqrt{2E})} \: {\rm exp}(-iEt)$
along a closed path, consisting of the positive real axis from $R$ to $0$, the negative 
imaginary axis from $0$ to $-iR$ and a quarter-circle, $\Gamma$, of radius $R$, is zero. 
The integral along $\Gamma$ is bounded by a constant times $1/R^k$ with $R = |E|$ and $k > 1$ 
and, consequently, vanishes in the limit $R \rightarrow \infty$. Then the integration over 
the real axis gives the same result as that over the imaginary axis. The variable change 
$E = -iy$ with $y$ real, then, yields
\begin{equation}
\Psi^{(I)}(x,t) = -i C_1 \int_{0}^{\infty} \phi(-iy) \: 
                  \sin{\left (x \: \sqrt{-2iy} \right )} e^{-yt} \: dy.
\end{equation}
For large times only small values of $y$ contribute to the integral.
The spectral function is substantially different from zero only close to the origin and can be
replaced by $\phi(0)$ and be pulled out of the integral while the sine function can be 
approximated by its argument in a finite range of $x$. The remaining integral is easily 
evaluated as a gamma function and gives
\begin{equation}
\Psi^{(I)}(x,t) \approx C_1 \phi(0) \: x \: e^{-i 3\pi /4} \sqrt{\pi /2} \: \: t^{-3/2}
\end{equation}  
confirming the earlier result. 

The survival probability, $P_{in}$, calculated thus far refers to the presence of the particle
inside the potential well. As it has been shown in the previous section the spectral function
of eq.~(\ref{spectralfunction}) produces non-zero probability density outside the well
at $t=0$ for $K > 0$. If the "interior" of the well is defined to extend to $x$ much larger
than $2L$ (without moving the delta function from $x = L$) then at $t=0$ the probability
to find the particle "inside" can be arbitrarily close to unity. Specifically the modified 
survival probability $P_{in}(4L)(t)$ can be defined by extending the integral of 
eq.~(\ref{survivalprob}) to $x = 4L$. This integral is then evaluated analytically and 
plotted in Fig. 4 as a function of time. As predicted and verified by an expansion of 
$P_{in}(4L)$ in inverse powers of time, its asymptotic time dependence is $t^{-3}$. An 
interesting feature of this plot is the presence of a step-wise behaviour which can be 
attributed to interference between waves moving in opposite directions.

\begin{figure}
\centerline{\psfig{figure=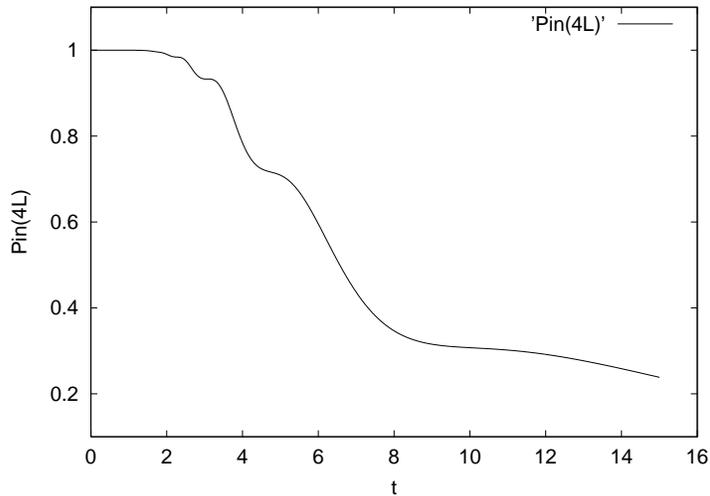,height=2.7in,angle=270}}
\caption{The modified survival probability for a potential consisting of an infinite wall and
a repulsive delta function and using a spectral function that is exponential in the energy
versus time (solid line). In this plot $L = 3$, $V_0 = 1$ and $K = 1/2$. The step-wise
behaviour is due to interference of waves moving in opposite directions.}
\label{deltaandwallpin4Lfig}
\end{figure}

\section{Conclusions and perspectives}

A method has been proposed to solve the time-dependent Schr\"{o}dinger equation utilizing
the much more easily obtained time-independent solutions for a given Hamiltonian. It has been
applied to the case of a potential consisting of an infinite wall and a repulsive delta function.
An exact, analytical, normalized solution has been obtained in closed form. The survival 
probability, which is also analytically calculated, exhibits a non-exponential behaviour.
At large times it decays approximately as $t^{-3}$. It was shown that this behaviour pertains
to all square-integrable wavefunctions that are solutions to this problem. To ensure
square-integrability the spectral function must be finite at $E \rightarrow 0$ and decrease 
to 0 at large energies faster than $1/\sqrt{E}$. With the appropriate 
choice of spectral functions which, due to linear independence need not be the same for waves 
propagating in different directions, the method could be applied to a variety of potentials. 
It is also of great interest to develop solutions for the time-dependent relativistic 
Dirac equation since these are more appropriate to describe the decay of mesons 
into light leptons. 

Acknowledgment:  Ames Laboratory is operated for the U.S. Department of Energy by 
Iowa State University under Contract No. W-7405-Eng-82.

\end{document}